\documentclass{pasj00}
\begin{document}
\SetRunningHead{S. Kato}{A Resonance Model of QPOs}
\Received{2005/02/27}%{yyyy/mm/dd}
\Accepted{2005/04/13}%{yyyy/mm/dd}

\title{A Resonance Model of Quasi-Periodic 
Oscillations of Low-Mass X-Ray Binaries}

%%% begin:list of authors
\author{Shoji \textsc{Kato}}
    %\thanks{Example: Present Address is xxxxxxxxxx}}
\affil{Department of Informatics, Nara Sangyo University, Ikoma-gun,
       Nara 636-8503}
\email{kato@io.nara-su.ac.jp, kato@kusastro.kyoto-u.ac.jp}

%\author{B-Firstname \textsc{B-Familyname}}
%\affil{B-Address of Institute}\email{bbbbb@xxx.xxx.xx.xx}
%\and
%\author{C-Firstname {\sc C-Familyname}}
%\affil{C-Address of Institute}\email{ccccc@xxx.xxx.xx.xx}
%%% end:list of authors

%%% Please use the following style in case that sorting by 
%%% affilation is impossible. 
%
% \author{%
%   D-Firstname \textsc{D-Familyname}\altaffilmark{1}
%   E-Firstname \textsc{E-Familyname}\altaffilmark{1,2}
%   and
%   F-Firstname \textsc{F-Familyname}\altaffilmark{2}}
% \altaffiltext{1}{Address of Institute}
% \email{ddddd@xxx.xxx.xx.xx}
% \email{eeeee@xxx.xxx.xx.xx}
% \altaffiltext{2}{Address of Institute}

%% `\KeyWords{}' always has to be placed before `\maketitle'.
\KeyWords{accretion, accrection disks 
          --- hectohertz QPOs
          --- horizontal branch QPOs
          --- kHz QPOs
          --- relativity
          --- resonance
          --- warp
          --- X-rays: stars} %Do NOT move this preamble from here!

\maketitle

\begin{abstract}

We try to understand the quasi-periodic oscillations (QPOs) in low-mass
neutron-star and black-hole X-ray binaries by a resonance model in 
warped disks with precession.
Our main concern is high-frequency QPOs,
hectohertz QPOs, and horizontal-branch QPOs in the z sources and the
atoll sources, and the correponding QPOs in black-hole X-ray binaries.
Our resonance model can qualitatively, but systematically, explain these
QPOs by regarding hectohertz QPOs as a precession of warp.

\end{abstract}

% \newpage

\section{Introduction}

Quasi-periodic oscillations (QPOs) have been observed
in many low-mass X-ray binaries.
They give important clues to understand disk structures as well as
to evaluate the mass and spin of the central neutron stars or black holes.
Although the mechanism of the QPOs is still under debate,
recent observations suggest that they can be attributed to disk oscillations.
Observations further suggest that some resonant processes are involved 
in the mechanism of the QPOs, and many disk oscillation models in this
 direction have been proposed since Abramowicz and Kluzniak (2001), e.g., 
Lamb and Miller (2003) and  Kluzniak et al. (2004).
In previous papers (Kato 2003, 2004a, 2004b) we have proposed that the 
QPOs are disk oscillations excited on a warped disk by a resonant process.
The purpose of this paper is to examine how much the warp model is
compatible with observations.

We consider a relativistic warped disk.
On the disk we superpose disk oscillations.
Then, some of these disk oscillations have resonant interactions 
with the disk 
at particular radii through non-linear coupling with the warp
(Kato 2003, 2004a, 2004b).
When the warp has no precession (this is the case considered in the
above papers) the resonances occur at radii
where one of the relations of $\kappa=(\sqrt{2}-1)\Omega$,
$\kappa=\Omega/2$, or $\kappa=(\sqrt{3}-1)\Omega$ is satisfied, where
$\kappa(r)$ is the epicyclic frequency and $\Omega(r)$ is the angular
frequency of disk rotation, $r$ being the radius from the disk center.
In the case of relativistic Keplerian disks, these radii are, in turn,  
$r=3.62r_{\rm g}$, $r=4.0r_{\rm g}$, and $r=6.46r_{\rm g}$ (Kato 2003,
2004a), where $r_{\rm g}$ is the Schwarzschild radius, defined 
by $r_{\rm g}=2GM/c^2$, $M$ being the mass of the central object.
Kato (2004b) subsequently showed that among the resonant oscillations
mentioned above, those at $\kappa=\Omega/2$ are excited spontaneously
by the resonance process, itself.
Kato (2004b) further showed that the high-frequency QPOs in black-hole
X-ray binaries, which are usually a pair and have a frequency ratio close
to 2 : 3, can be explained by this warped model.
 
In this paper we extend the warped disk model to the case where
the warp has precession.
We demonstrate that by this extension the main important characteristics 
of QPOs in X-ray binaries (neutron-star and black-hole X-ray binaries) can 
be qualitatively explained.
Among three types of resonances, which tend, in the limit of no precession, 
to $\kappa=(\sqrt{2}-1)\Omega$,
$\kappa=\Omega/2$, or $\kappa=(\sqrt{3}-1)\Omega$, we focus our attention 
in this paper on the middle one, since the resonances 
in this case spontaneously excite oscillations, as mentioned above.

\section{Resonant Oscillations at $\kappa=(\Omega+\omega_{\rm p})/2$ 
on Warped Disks}

Details of the resonance process on warped disks are
presented by Kato (2003, 2004a, 2004b) in the case where 
the warp has no precession.
The essence of the resonance processes is the same, even when a warp has 
precession.
We thus present here only an outline. 
An overview of our non-linear resonance model in the case where the
warp has precession is sketched in figure 1 of Kato (2004c).

We consider geometrically thin disks rotating with angular velocity
$\Omega(r)$.
The epicyclic frequency on the disk is denoted by $\kappa(r)$.
The oscillations on  geometrically thin disks are generally classified 
into g-mode and p-mode oscillations (see, e.g., Kato et al. 1998; Kato 2001).
In simplified disks the oscillations are further classified by the 
set of ($m$, $n$), where
$m=$(0, 1, 2...) is the number of nodes in the azimuthal direction, and 
$n=$(0, 1, 2...) is a number related to nodes in the vertical direction.
That is, $n$ represents the number of nodes that $u_r$ (the radial 
component of velocity associated with oscillations) has in
the vertical direction.
It is noted, however, that $u_z$ (the vertical component of velocity
associated with oscillations) has ($n-1$) nodes in the vertical direction,
and $u_z=0$ in the case of $n=0$.
    
A warp is a global deformation of disks with $m=n=1$.
The warp is assumed to have a precession whose angular frequency is 
$\omega_{\rm p}$,
On a disk deformed by the warp we superpose g-mode oscillations with
arbitrary $m$ and $n$.
A g-mode oscillation with frequency $\omega$ and ($m$, $n$) has 
a relatively large amplitude, global pattern only around the radius where
\begin{equation}
    (\omega-m\Omega)^2-\kappa^2 = 0
\label{1.1}
\end{equation}
is satisfied.
This can be understood if the dispersion 
relation for local perturbations is considered (e.g., Kato et al. 1998;
Kato 2001). 
That is, the region of $(\omega-m\Omega)^2-\kappa^2 >0$ is a
evanescent region of the oscillations.
In the region where 
$(\omega-m\Omega)^2$ is smaller than $\kappa^2$, on the other hand,
the oscillations have
very short wavelengths in the radial direction in geometrically thin
disks.   

A non-linear interaction of this g-mode oscillation with the warp
produces an oscillation with $\omega\pm\omega_{\rm p}$, ${\tilde m}$ 
and ${\tilde n}$, where ${\tilde m}=m\pm 1$ and ${\tilde n}=n\pm 1$
(these oscillations are called hereafter intermediate oscillations).
These intermediate oscillations resonantly interact with the disk at 
the radius where 
the dispersion relation for these intermediate oscillations is 
satisfied [see Kato (2004b) for detailed discussions].
There are two types of resonances.
One is resonances that occur through motions in the vertical direction, 
and the other is those through motions in the radial direction (see
Kato 2004a, referred to Paper I).
Here, we are interested in resonances in the horizontal direction. 
The horizontal resonances occur around the radius where 
\begin{equation}
   (\omega\pm \omega_{\rm p}-{\tilde m}\Omega)^2-\kappa^2 \sim 0
\label{1.2}
\end{equation}
is satisfied.
Combining equations (\ref{1.1}) and (\ref{1.2}), we find that the resonances 
occur at the radius of $\kappa=\Omega/2\pm \omega_{\rm p}/2$ 
(cf. Paper I).
After this resonance the intermediate oscillations feedback to the
original oscillations, amplifying or dampening the original oscillations
(Kato 2004b).
Hereafter, we consider the case of $\kappa=\Omega/2+\omega_{\rm p}/2$.

A detailed examination shows that when ${\tilde m}$ of the
intermediate oscillation is $m-1$, i.e., ${\tilde m}=m-1$,
the oscillations that resonantly
interact with the disk at $\kappa=(\Omega+\omega_{\rm p})/2$ are 
those satisfying $\omega=m\Omega-\kappa$ at the resonant radius.
On the other hand, the oscillations that resonantly interact with the
disk at $\kappa=(\Omega+\omega_{\rm p})/2$ are those satisfying
$\omega=m\Omega+\kappa$ there, when ${\tilde m}=m+1$ (see Paper I).

Here, we consider non-axisymmetric oscillations.
Among them we are particularly interested in oscillations of
a small number of $m$.
Typical ones are $\omega=\Omega-\kappa$ ($m=1$, ${\tilde m}=0$),
$\omega=\Omega+\kappa$ ($m=1$, ${\tilde m}=2$),
and $\omega=2\Omega-\kappa$ ($m=2$, ${\tilde m}=1$).
As shown below, we identify these oscillations, in turn, to the
horizontal branch QPOs, upper-frequency kHz QOPs, and lower-frequency
kHz QPOs in the case of z sources.
Considering this, we introduce the following notations:
\begin{equation}
  \omega_{\rm H}=\Omega+\kappa, \quad \omega_{\rm L}=2\Omega-\kappa,
      \quad  \omega_{\rm HBO}=\Omega-\kappa.
\label{1.3}
\end{equation}

\section{Precession of Warps}

We consider a relativistic Keplerian disk with the Schwarzschild
metric, and examine the radii where the resonance condition, 
$\kappa=(\Omega+\omega_{\rm p})/2$, is satisfied as functions of
$\omega_{\rm p}$.
The condition is satisfied at two
different radii when $\omega_{\rm p}>0$.
(In the case of $\omega_{\rm p}=0$, we have only one radius, i.e., 
$4.0\ r_{\rm g}$.
The other one is $\infty$.)
As $\omega_{\rm p}$ increases, the inner radius becomes larger than 
$4.0\ r_{\rm g}$, while the other one decreases from infinity.
At a certain critical value of $\omega_{\rm p}$, both radii
coincide and above the critical value of $\omega_{\rm p}$, there is no
solution of $\kappa=(\Omega+\omega_{\rm p})/2$.
The results are shown in figure 1.
The unit of the abscissa is $\omega_{\rm p}(M/M_\odot)$. 
The critical value of $\omega_{\rm p}$ is $\sim 325$ Hz when $M/M_\odot=1.0$,
while $\sim 162$ Hz  when $M/M_\odot=2.0$.
The branch of larger value of $r$ on the $r$--$\omega_{\rm p}$ plane 
is hereafter called 
the upper branch and that of the lower one the lower branch.
The frequencies ($\omega_{\rm H}$, $\omega_{\rm L}$, $\omega_{\rm HBO}$)
at radii of the lower branch decrease as $\omega_{\rm p}$ increases, 
since the resonance radius moves outward.
On the other hand, the frequencies at radii of the upper branch increase as
$\omega_{\rm p}$ increases.
To compare with observations, the $\omega_{\rm L}$--$\omega_{\rm H}$,
$\omega_{\rm HBO}$--$\omega_{\rm H}$ and 
$\omega_{\rm p}$--$\omega_{\rm H}$ relations are shown in figure 2,
which is free from precession.  
For a comparison, the straight line of the $\omega_{\rm H}$--$\omega_{\rm H}$
relation is also added.
Values of $\omega_{\rm p}(M/M_\odot)$  
are shown, for convenience,  at some points on the 
$\omega_{\rm p}$--$\omega_{\rm H}$ curve..
The trancated curve on the upper-right corner is the first harmonic of
$\omega_{\rm HBO}$, i.e., it represents the 
$2\omega_{\rm HBO}$--$\omega_{\rm H}$ relation.
The curve extends  to the left, but is cutted in order to avoid 
complexity of the figure.
It is useful to compare this figure with figure 2.9 of van der Klis (2004).
We take the standpoint that the variation of the precession is the cause 
of time variations of QPO frequencies in a single object.

As the precession frequency changes, the frequencies
$\omega_{\rm H}$, $\omega_{\rm L}$, and $\omega_{\rm HBO}$ vary along 
the curves in figure 2.
Observations show that the changes of the horizontal branch QPOs and 
lower-frequency kHz QPOs are correlated so that   
the former frequencies are $\sim$ 0.08-times the latter ones 
(Psaltis et al. 1999).
Hence, for a comparison, the curve of the 
$0.08\ \omega_{\rm L}$--$\omega_{\rm H}$ relation is shown in figure 2.
Figure 2 shows that the curve of the 
$0.08\ \omega_{\rm L}$--$\omega_{\rm H}$ relation 
crosses the curve of the $\omega_{\rm HBO}$--$\omega_{\rm H}$ relation
at $\omega_{\rm H}\sim 320(M/M_\odot)^{-1}$ Hz, which corresponds to
$\omega_{\rm p}\sim 115(M/M_\odot)^{-1}$ Hz.

It is noted that the curves shown in figure 2 are mass-independent, 
since the axes are normalized by $M/M_\odot$.
That is, the results hold in a wide range of frequency by
changing the mass.

\section{Summary and Discussion}

In luminous neutron-star low-mass X-ray binaries (the z sources),
we typically have four distinct types of QPOs.
These are the $\sim 5$--20 Hz normal branch oscillation (NBO),
the 15--60 Hz horizontal branch oscillation (HBO), and the
$\sim 200$--1200 Hz kilohertz QPOs that typically occur in pairs.
The QPOs in the atoll sources (less luminous neutron-star low-mass X-ray 
binaries) can also be classified into similar types of oscillations.
In addition, in the atoll sources the hectohertz QPOs are observed in 
the frequency range of 100 Hz--200 Hz. 
In the z sources, however, the presence of hectohertz QPOs is uncertain,
or ambiguous.   
In black-hole X-ray binaries, we have  
high-frequency QPOs in the range of 100 Hz to 450 Hz,
which usually appear in pairs.
A comprehensive review on QPOs is presented by van der Klis (2004).

One of important characteristics of QPOs in the z sources is the presence of 
a strong correlation between kilohertz QPOs and HBOs.
That is, the frequency of the lower 
kHz QPO and that of HBO are correlated
in each object so that the former is $\sim 0.08$-times the latter.
This correlation extends from
neutron-star to black-hole X-ray binaries in nearly three
orders of magnitude in frequency (Psaltis et al. 1999, 
see figure 2 of their paper).
In our resonance model this correlation can be explained if
variation of precession frequency occurs around 115$(M/M_\odot)^{-1}$ Hz,
say, $70(M/M_\odot)^{-1}$ Hz $\sim$ $170(M/M_\odot)^{-1}$ Hz.
A question is whether such a precession has been observed.
In relation to this issue, it it interesting to note that the observed 
hectohertz QPOs are roughly in the frequency range mentioned above.
This suggests that the hectohertz QPOs might be a manifestation 
of the warp.
One of the characteristics of the hectohertz QPOs is that their
frequencies are nearly constant (e.g., van der Klis 2004).
In our resonance model, the relation of 
$\omega_{\rm HBO}\sim 0.08\ \omega_{\rm L}$ is
realized over in a wide range of $\omega_{\rm H}$ without much changing 
the value of $\omega_{\rm p}$, as shown in figure 2.
This is consistent with the idea that the warp represents the 
hectohertz QPOs.
The precession of disks in X-ray binaries 
is theoretically expected, since the radiation force from central star
gives
torques on warped disks (Pringle 1996; Maloney et al. 1996).
It is not clear, however, whether such a high-frequency precession as 
required here is generally expected.
We suppose that the frequency of the precession is related to  rotation
of the central star.
A non-axisymmetric pattern on a rotating stellar surface will give rise
to a precession of the disk through the effects of a radiative force or 
a magnetic field.

Another important characteristic of QPOs in neutron-star X-ray binaries
is that the frequency ratio of the pair kHz QPOs is close to 3 : 2,
but changes with time so that the ratio decreases with an increase of
the frequency.
This observational trend is realized in our model.
In our model the ratio $\omega_{\rm H}/\omega_{\rm L}$ is 1.5 at
$\omega_{\rm H}\sim 700(M/M_\odot)^{-1}$ Hz, and becomes larger than 1.5
for smaller $\omega_{\rm H}$.
For larger $\omega_{\rm H}$ the ratio tends to unity as the resonant
radii approaches to $4.0$ $r_{\rm g}$ (i.e., $\omega_{\rm p}=0$).

Next, let us consider black-hole X-ray binaries.
The pair of high-frequency QPOs in these objects changes little their 
frequencies, keeping the ratio close to 3 : 2.
This is different from the case of neutron-star 
X-ray binaries.
Our resonance model qualitatively explains this.
Considering their mass and the observed frequencies of QPOs, 
we think that in the case of black-hole X-ray binaries the observed 
QPOs are those resulting from the lower branch of the $r$--$\omega_{\rm p}$
relation (see figure 1).
That is, the resonance radius is close to 4.0 $r_{\rm g}$.
This case corresponds to the upper-right corner of figure 2.
In the limit of $\omega_{\rm p}=0$, the resonance occurs at $r=4.0\ r_{\rm 
g}$ and the frequency ratio of $\omega_{\rm L}$ and the first harmonic of
$\omega_{\rm HBO}$ is just 3 : 2.
(It is noted that the first harmonic of $\omega_{\rm HBO}$ has been 
also observed in the z sources.)
These QPOs change little their frequencies for a
change of precession frequency, since as shown in figure 1 the
resonance radius remains close to $4.0\ r_{\rm g}$ for a large change of 
$\omega_{\rm p}$.
Another explanation of little change of frequencies of high-frequency
QPOs in black-hole binaries is that the precession is really 
small in the case of black holes, since  
the radiation force from the central object is absent.
It is noted that in our model the modes of the pair QPOs in the black-hole
binaries are different from those in neutron-star binaries.
That is, the pairs in the former are $\omega_{\rm H}(=\omega_{\rm L})$ and
$2\omega_{\rm HBO}$, while those in the latter are $\omega_{\rm H}$
and $\omega_{\rm L}$.
 
We have restricted our attention only to  
the resonances at $\kappa=(\Omega+\omega_{\rm p})/2$,
since the analyses of growth rate of resonant oscillations suggest 
that the resonances at $\kappa=(\sqrt{2}-1)\Omega+\omega_{\rm p}$
and $\kappa=(\sqrt{3}-1)\Omega+\omega_{\rm p}$ are not spontaneously excited
(Kato 2004b).
However, examinations of these cases are worthwhile.
As an example, some results in the case of 
$\kappa=(\sqrt{2}-1)\Omega+\omega_{\rm p}$ are shown in figure 3.
For simplicity, in this figure, only the resonant oscillations resulting 
from the upper branch of the $r$--$\omega_{\rm p}$ curve are shown.
The frequency of precession required to obtain 
$\omega_{\rm HBO}\sim 0.08\ \omega_{\rm L}$ is 
around 50$(M/M_\odot)^{-1}$ Hz--100$(M/M_\odot)^{-1}$ Hz.
 
Our model predicts that some QPOs that are not yet observed are 
present in neutron-star and black-hole X-ray binaries.
In our model the observed QPOs in neutron-star X-ray binaries are the
resonance oscillations at resonant radii belonging to the upper
branch of the $r$--$\omega_{\rm p}$ diagram (figure 1).
Resonances coming from a resonance radius of the lower branch are
also expected.
They have higher frequencies compared with the observed ones.
The observed QPOs in black-hole X-ray binaries, on the other hand,
are interpreted to be resonant oscillations belonging to  
the lower branch of the $r$--$\omega_{\rm p}$ relation.
Resonant oscillations resulting from  radii of the
upper branch are also expected.
These oscillations have lower frequencies compared with the observed
ones.
 
\bigskip\noindent
A note added on April 30:

In the present paper we have considered horizontal resonances
of g-mode oscillations.
In this model, the precession required is retrograde.
Furthermore, time variation of the precession is required 
in order to explain time variation of QPO frequencies.
In the case of vertical resonances of g-mode oscillations, 
however, the observed QPO frequencies and their time variations
can be explained
as a result of time change of disk structure in the
vertical direction, without appealing to precession.
This is discussed in a subsequent paper.

%It is said that the frequency correlation between the horizontal branch 
%QPOs and the kHz QPOs can be extended to cataclysmic variables (
%Mauche 2002; Warner, Woudt 2005).
%Our model can potentially explain such a  
%frequency correlations covering a wide range of frequency.
%In the case of white dwarfs the disk is Newtonian, but
%the resonance condition can be satisfied by the disk rotation
%being modified from the Keplerian one  by pressure force.
%Hence, it is not suprising that the correlation extends from a
%relativistic disk to a non-relativistic disk.
%This issue will be discussed in the future.

\bigskip
\leftskip=20pt
\parindent=-20pt
\par
{\bf References}
\par
Abramowicz, M. A., \& Kluzniak, W. 2001, A\&A, 374, L19 \par
Kato, S. 2001, PASJ, 53, 1\par 
Kato, S. 2003, PASJ, 55, 801\par
Kato, S. 2004a, PASJ, 56, 559 (Paper I)\par
Kato, S. 2004b, PASJ, 56, 905\par
Kato, S. 2004c, PASJ, 56, L25 \par
Kato, S., Fukue, J., \& Mineshige, S. 1998, Black-Hole Accretion Disks 
  (Kyoto: Kyoto University Press) ch. 13 \par
Kluzniak, W., Abramowicz, M. A., Kato, S., Lee, W. H., \& Stergioulas,
   N. 2004, ApJ, 603, L89 \par 
%Kluzniak, W. \& Abramowicz, M. 2003?,   \par
Lamb, F. K., \& Miller, M. C. 2003, astro-ph/0308179 \par
Maloney, P. R., Begelman, M. C., \& Pringle, J. E. 1996, ApJ, 472, 582\par
%Mauche, C. W. 2002, ApJ, 580, 423\par 
Pringle, J. E. 1992, MNRAS, 258, 811\par  
Psaltis, D., Belloni, T., \& van der Klis, M. 1999, ApJ, 520, 262\par
van der Klis, M. 2004, astro-ph/0410551 \par 
%Warner, B.,\& Woudt, P.A. 2004, in The Astrophysics of Cataclysmic Variables
%and Related Objects, ed. J. M. Hameury \& P.-J. Lasota (ASP Conf.
%   Ser.), astro-ph/0409287 \par
\bigskip\bigskip

\begin{figure}
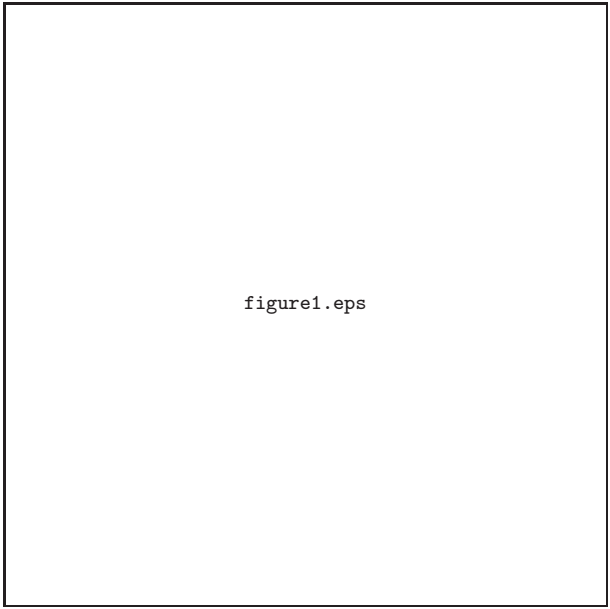

  \begin{center}
    \FigureFile(80mm,80mm){figure1.eps}
    %%% \FigureFile(width,height){filename}
  \end{center}
  \caption{
The $r$--$\omega_{\rm p}$ relation giving the solution of 
the resonance condition $\kappa=(\Omega+\omega_{\rm p})/2$.
The disk is Keplerian in the Schwarzschild metric.}
\label{fig:sample}
\end{figure}

\begin{figure}
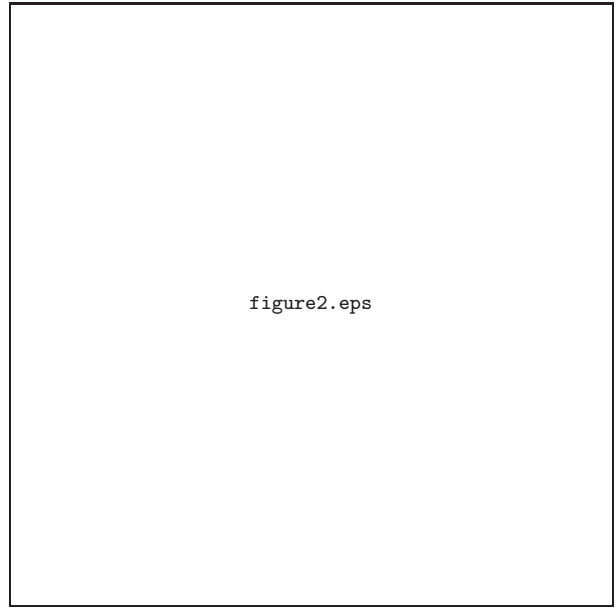

  \begin{center}
    \FigureFile(80mm,80mm){figure2.eps}
    %%% \FigureFile(width,height){filename}
  \end{center}
  \caption{
The $\omega_{\rm L}$--$\omega_{\rm H}$, 
$\omega_{\rm HBO}$--$\omega_{\rm H}$, and 
$\omega_{\rm p}$--$\omega_{\rm H}$ relations
obtained from the resonance condition $\kappa=(\Omega+\omega_{\rm p})/2$ 
for some frequency range of $\omega_{\rm H}$.
For a comparison, the line of $\omega_{\rm H}$--$\omega_{\rm H}$ is shown 
and the value of $\omega_{\rm p}$ along the 
$\omega_{\rm p}$--$\omega_{\rm H}$ curve is shown at some points.
For a comparison, the 
$0.08\ \omega_{\rm L}$--$\omega_{\rm H}$ relation is shown.
The trancated curve on the upper-right corner represents the first
 hamonic of $\omega_{\rm HBO}$, i.e., it is the 
$2\omega_{\rm HBO}$--$\omega_{\rm H}$ relation.
In order to avoid complexity, the curve is trancated.
It is noted that the ratio $\omega_{\rm H}/\omega_{\rm L}$ decreases
as $\omega_{\rm H}$ increases.
The ratio is $\sim 1.5$ for $\omega_{\rm H}$ in the middle of the figure, 
and becomes 1.0 at the right end of the figure, where $\omega_{\rm H}$ 
is the maximum and $\omega_{\rm p}=0$.
At this right end, the ratio $\omega_{\rm H}(=\omega_{\rm L})$ : 
$2\omega_{\rm HBO}$ : $\omega_{\rm HBO} =$ 3 : 2 : 1.
We regard the oscillations of $\omega_{\rm p}$ as hectohertz QPOs.
}
\label{fig:sample}
\end{figure}

\begin{figure}
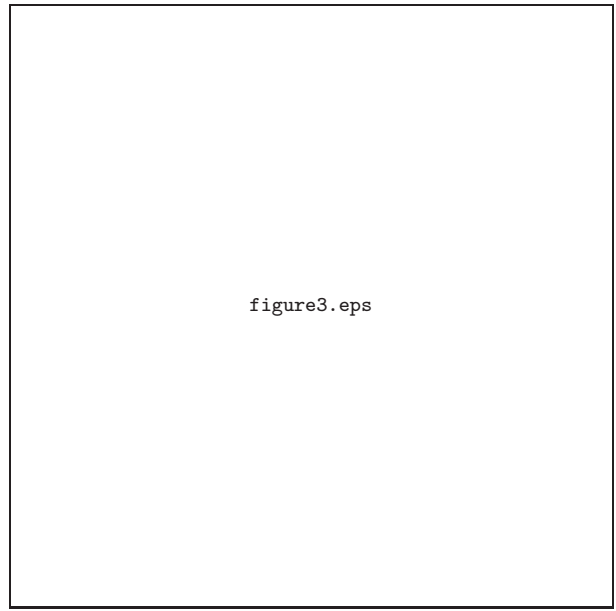

  \begin{center}
    \FigureFile(80mm,80mm){figure3.eps}
    %%% \FigureFile(width,height){filename}
  \end{center}
  \caption{
Same as figure 2, except for $\kappa=(\sqrt{2}-1)\Omega+\omega_{\rm p}$.
In this figure, for simplicity, only the resonant oscillations that 
 occur on the upper branch of the $r$--$\omega_{\rm p}$ plane are
 shown.}
\end{figure}

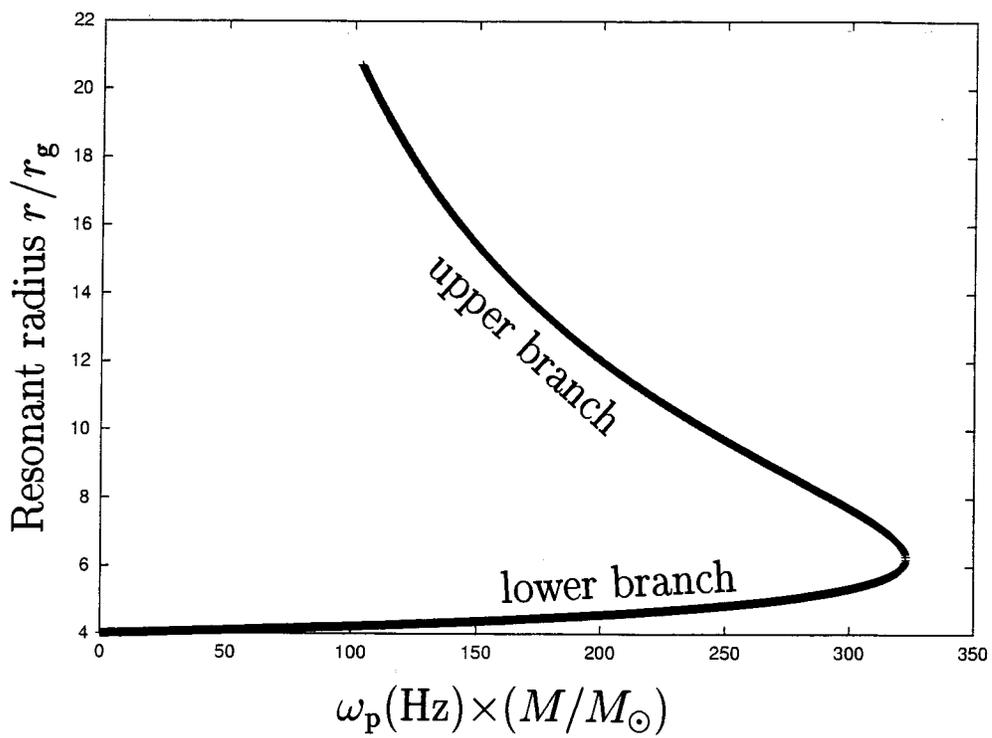
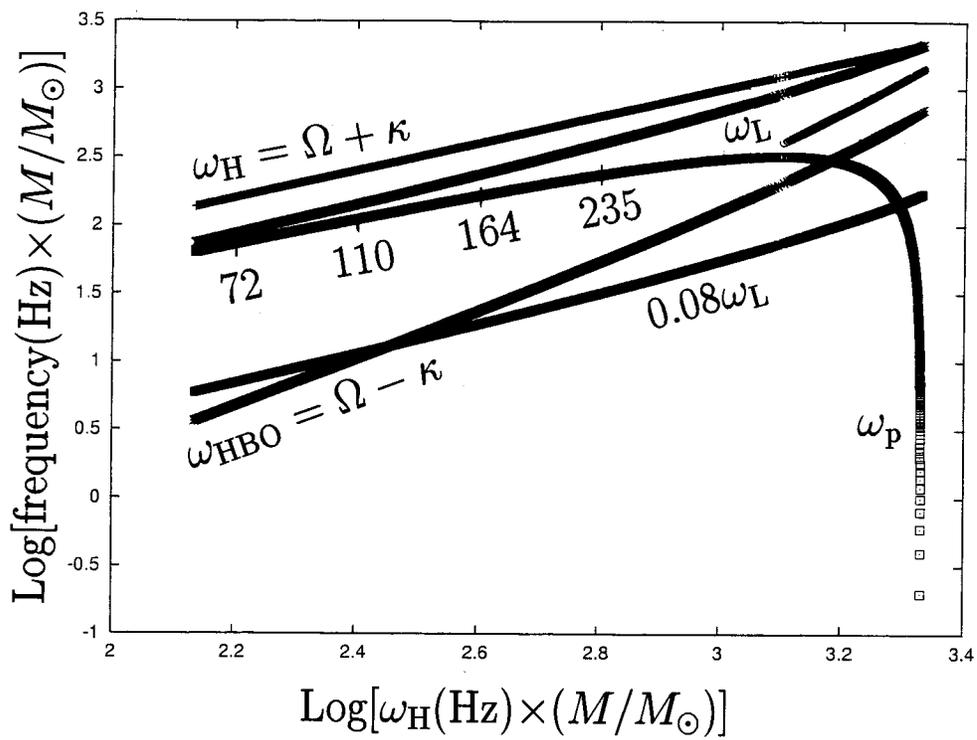
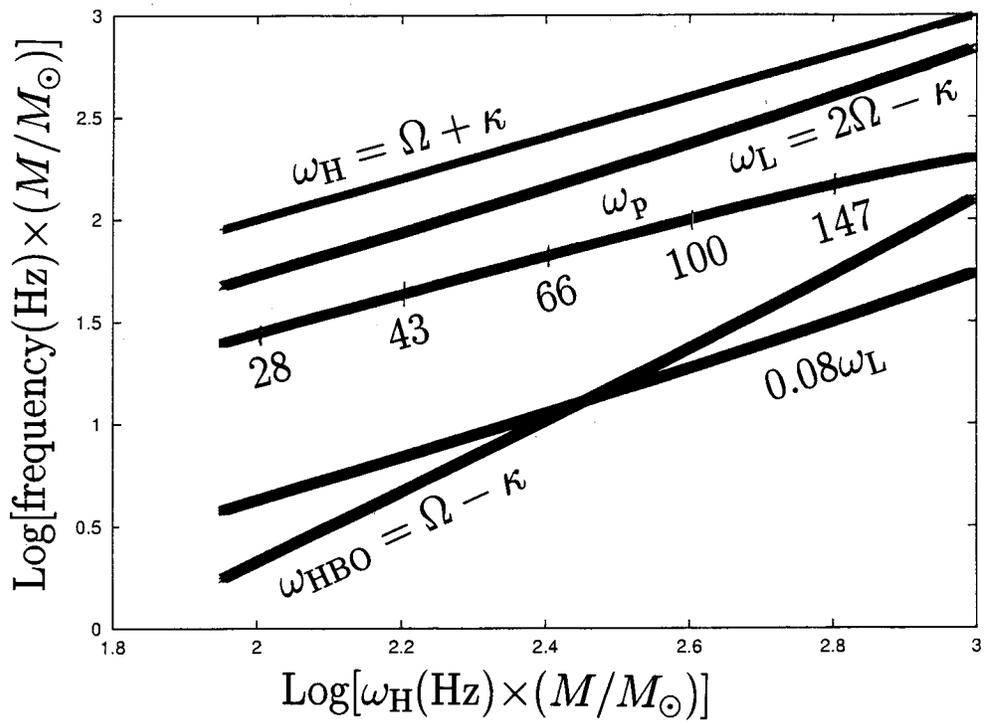
\end{document}